\documentclass[12pt,preprint]{revtex4}
\usepackage{graphicx}
\usepackage{xspace}

\def\Vec#1{\mbox{\boldmath $#1$}}
\def\beq{\begin{equation}}
\def\eeq{\end{equation}}

\def\beqy{\begin{eqnarray}}
\def\eeqy{\end{eqnarray}}

\newcommand{\herwig}{\mbox{\textsc{Herwig}}\xspace}
\newcommand{\pythia}{\textsc{Pythia}\xspace}
\newcommand{\pt}{$p_{\rm T}$\xspace}
\newcommand{\ptp}{p_{\rm T}} 
\newcommand{\sigmaeff}{$\sigma_{\mathrm{eff}}$}
\newcommand{\sigmaeffp}{\sigma_{\mathrm{eff}}} 

\usepackage{graphics}
\begin{document}
\title{Revealing minijet dynamics via centrality dependence of double parton interactions in proton--nucleus collisions}
%
\author{M. Alvioli}
\affiliation{Consiglio Nazionale delle Ricerche, Istituto di Ricerca per la 
  Protezione Idrogeologica, via Madonna Alta 126, I-06128 Perugia, Italy}
\author{M. Azarkin}
\affiliation{Lebedev Physics Institute, Moscow 119991, Russia}
\author{B. Blok}
\affiliation{Department of Physics, Technion -- Israel Institute of Technology,
  Haifa, 32000 Israel}
\author{M. Strikman}
\affiliation{104 Davey Lab, The Pennsylvania State University,
  University Park, PA 16803, USA}
%
%
%
\begin{abstract}
  One of the main challenges hampering an accurate measurement of the double parton scattering (DPS)
  cross sections is the difficulty in separating the DPS from the leading twist (LT) contributions.
  We argue that such a separation can be achieved, and cross section of DPS measured, by exploiting
  the different centrality dependence of DPS and LT processes in proton--nucleus scattering.
  We developed a Monte Carlo implementation of the DPS processes which includes realistic nucleon--nucleon
  (NN) correlations in nuclei,   an accurate description of transverse geometry of both hard and soft
  NN collisions as well as fluctuations of the strength of interaction of nucleon with nucleus (color fluctuation effects).
  Our method allows the calculation of probability distributions of single and double dijet
  events as a function of centrality, also distinguishing double hard scatterings originating from a single
  target nucleon and from two different nucleons.
  We present numerical results for the rate of DPS as a function of centrality, which relates the number
  of wounded nucleons and the sum of transverse energy of hadrons produced at large negative (along the
  nucleus direction) rapidities, which is experimentally measurable.
  We suggest a new quantity which allows to test the geometry of DPS and we argue that it is a universal
  function of centrality for different DPS processes.
  This quantity can be tested by analyzing existing LHC data. The method developed in this work can be
  extended to the search for triple parton interactions.
\end{abstract}
\maketitle
\thispagestyle{empty}
%
%
\section{Introduction}

At the LHC energies a typical proton--proton (\textit{pp}) collision involves several parton-parton interactions
with transverse momentum transfer of a few GeV, leading to the production of several minijets, which are referred
to as multiparton interactions (MPI). Successful Monte Carlo (MC) models of \textit{pp} inelastic interaction at
the LHC, such as the models implemented in the event generators \pythia \cite{Sjostrand:2014zea} and \herwig
\cite{Bellm:2015jjp}, have to tame such parton-parton interactions up to \pt $\sim 4$~GeV.
Within these models, the taming has to strengthen with the increase of the invariant energy of collision. Minijets
give an important contribution to the production of relatively soft hadrons that give a main contribution to the so
called underlying event (UE) with respect to the hard processes. It is generally accepted that characteristics of
the UE are measured in the direction perpendicular to the momentum of a high-\pt jet \cite{Cacciari:2009dp}.
However, a direct observation of minijets is challenging since it is very difficult to separate them. Over
the last decade intensive theoretical and experimental studies of double parton scattering (DPS) were performed
\cite{Blok:2010ge,Diehl:2011tt,Blok:2011bu,Diehl:2011yj,Blok:2013bpa}; a comprehensive reviews was recently
compiled in Ref. \cite{Bartalini:2017jkk}. 
  
\begin{figure}[!htp]
  \centerline{{\includegraphics[width=0.5\textwidth]{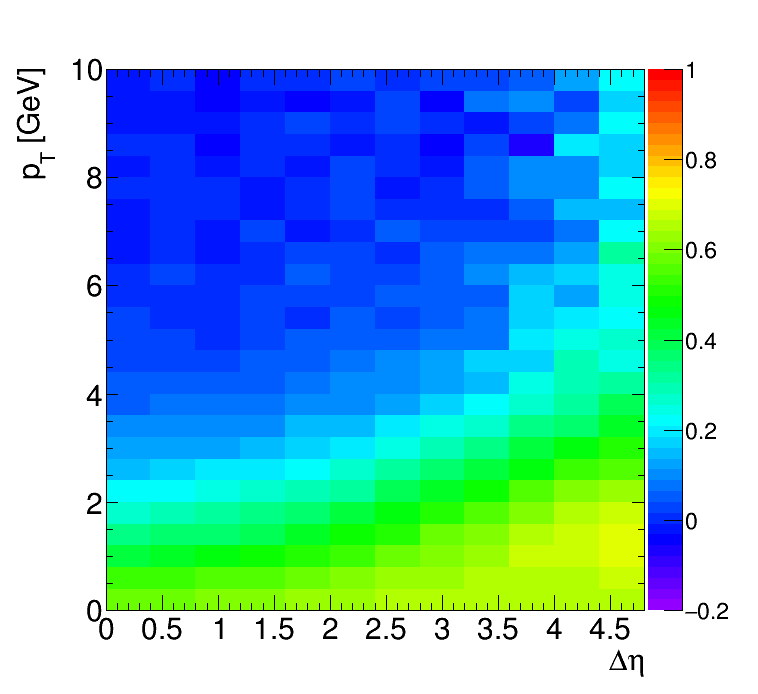}}}
  \caption{Fraction of dijet + a charged particle cross section due to the DPS as
    a function of the \pt of the charged particle and the pseudorapidity interval
    between the leading jet and the charged particle, $\Delta \eta$. The result
    is obtained using \pythia 8 Monash model \cite{Skands:2014pea}.}
  \label{LTvsHT}
\end{figure}

In particular, a number of experimental analyses have been performed, aiming at finding an optimal kinematics
where the ratio of the cross sections of DPS to the competing leading twist processes are somewhat enhanced.
Except the case of the double charm production \cite{Belyaev:2017sws,Likhoded:2015zna,Blok:2016lmd}, the best
kinematics still corresponds to the DPS being a correction to the LT contribution. Hence, the identification of
DPS events is rather sensitive to the particular model adopted to describe LT processes, which are usually rather
involved. To illustrate this point, Fig. \ref{LTvsHT} shows the fraction of the total cross section of dijet
production within $\vert\eta \vert <$~2 and $\ptp^{\rm jet} \ge$ 50 GeV plus a charged particle, which originates
from the different parton interaction, with an azimuthal angle difference with respect to the leading jet within
$80^o < \Delta \phi < 100^o$, as a function of the pseudorapidity interval between the leading jet and the charged
particle, $\Delta \eta$, and of the transverse momentum, $\ptp$, of the charged particle as obtained with \pythia 8
Monash model \cite{Skands:2014pea}. The fraction of the cross section due to DPS presented in Fig. \ref{LTvsHT} is
computed as a difference between the standard collision simulation and one with MPI mechanism switched-off, divided
by former one. One can see from Fig. \ref{LTvsHT} that the DPS contribution is significant but not dominant, hence
a relatively small uncertainty in the calculation of the LT contribution leads to a pretty large uncertainty in
the determination of the DPS contribution to the experimental cross section. 
Traditionally the DPS cross section is parameterized in the following form:
\begin{equation}
  \sigma^{\rm DPS}\,=\,\frac{\sigma_1\,\sigma_2}{\sigmaeffp}\,,
  \label{sigmaeff}
\end{equation}
where $\sigma_i$ are the cross sections of binary \textit{pp} collisions, and \sigmaeff~is the
effective cross section, widely used to characterize the effective transverse area of hard partonic
interactions in \textit{pp} collisions \cite{Paver:1984ux,Mekhfi:1983az}.

In QCD one expects \sigmaeff~to depend on the Bjorken $x$'s of the colliding partons, their
flavors, as well as the hardnesses of the subprocesses. We will not write this dependence explicitly
in the following.  
 
The LHC data are consistent with $\sigmaeffp \sim$ 20 mb for production of two pairs of jets with
$\ptp^{\rm jet} \ge$ 50 GeV \cite{Aaboud:2016dea}. In this paper we use the formalism for the description
of MPI developed in \cite{Blok:2010ge,Blok:2011bu,Blok:2013bpa,Blok:2016lmd}, see review and references
in \cite{Blok:2017alw}, which takes into account both the mean field contributions as well pQCD--induced
parton--parton correlations and small $x$ soft correlations. This formalism allows to describe all
existing LHC data except double $J/\psi$ production \cite{Aaboud:2016fzt}. For smaller virtualities
this formalism predicts $\sigmaeffp \sim$ 30 mb, which is consistent with the recent Monte Carlo analyses
\cite{Blok:2015rka}. The model also explains an increase of $\sigmaeffp$ from $\sim$ 14 mb to 20 mb between
the Tevatron and LHC energies for the kinematical ranges in which measurements were performed.
 
Though the LHC data strongly suggest the presence of the MPI effects in \textit{pp} scattering, no accurate
determinations of the MPI cross were reported so far (a notable exception is the charm production
\cite{Belyaev:2017sws,Likhoded:2015zna,Blok:2016lmd}). To a large extent, this is due to insufficient
accuracy of modeling higher order leading twist (LT) contributions to multijet production.
  
We suggest that a way out is to study MPI in proton--nucleus collisions as a function of the centrality
of the collision. The suggested procedure is based on the observation made a long time ago \cite{Strikman:2001gz}
that MPI are enhanced in proton--nucleus collisions, leading to a parametric enhancement of MPI by the factor
$\propto A^{1/3}$ as compared to the LT contribution due to hard scattering off two nucleons. 
The enhancement strongly increases with centrality of the collision. Hence, the study of the rate of the MPI
candidate events as a function of centrality would allow to separate DPS and LT
processes and provide an unambiguous measurement of DPS.

We study the centrality dependence of the different contributions to DPS in $pA$ collisions at LHC energies,
within a high-accuracy implementation of the Glauber Monte Carlo model. Our model makes use of realistic
nucleus configurations including NN correlations \cite{Alvioli:2009ab} and neutron skin in the lead nucleus
\cite{Alvioli:2018jls}. Other implementations of the Monte Carlo Glauber model for soft processes exist, for
example the one in Ref. \cite{Loizides:2017ack}. In the treatment of the individual soft $pN$ collisions, we
also include the color fluctuation effect \cite{Alvioli:2014eda}, which takes into account
the possibility for the incoming proton to fluctuate in different quantum states with substantially different
$pN$ interaction strength; this effect 
is important for an accurate description of the dependence of the hadron production on centrality  \cite {Aad:2016zif},
see discussion in Sec.\ref{sec:sec3}. The main effect of smearing of centrality
which we take into account is due to the experimental definition of centrality classes, based on the measured
transverse energy distribution $\sum E_{\rm T}$. Eventually, we implement a mechanism for a double hard trigger
in each Monte Carlo Glauber event, based on the extension to two hard interactions of an existing model for
single hard trigger \cite{Alvioli:2014sba}.

We organized the paper as follows.

\noindent 
In section \ref{sec:sec0} we describe the basic idea and summarize the relevant information from the previous
studies. Next, in section \ref{sec:sec2} we describe the development of a Monte Carlo event generator for
calculating the inclusive rate of DPS. In section \ref{sec:sec3} we describe an extension to the case of
DPS of the existing Monte Carlo procedure for the calculation of the probability distribution over the number
of the wounded nucleons in events with single hard interaction. In section \ref{sec:sec4} we include the effect
of smearing over impact parameter for the transverse energy of hadrons for centrality characterization. Based
on this calculation we outline the proposed procedure for comparing events of different centrality classes in
order to measure the DPS cross section.

\section{Basic idea}
\label{sec:sec0}

\begin{figure*}[!htp]
  \begin{minipage}[h]{0.49\linewidth}
    \center{\includegraphics[width=1\linewidth]{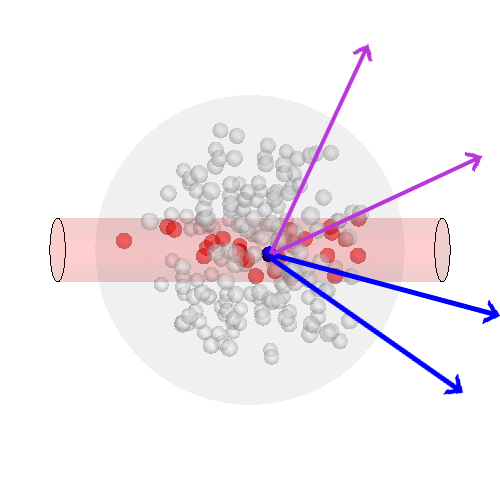} \\ \vskip -1.0cm {\Large(a)}}
  \end{minipage}
  \begin{minipage}[h]{0.49\linewidth}
    \center{\includegraphics[width=1\linewidth]{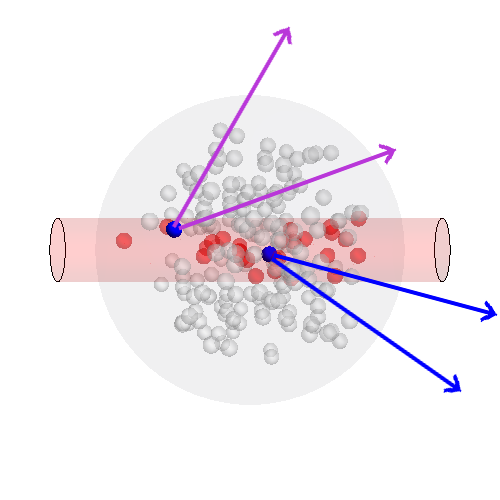} \\ \vskip -1.0cm {\Large(b)}}
  \end{minipage}
  \vskip 0.2cm
  \caption{Sketch of double parton collisions with production of four jets (arrows on the plot)
    occurring on a single nucleon (a) or on two different nucleons (b) in the target nucleus.
    In both illustrations, hard-interacting nucleons are depicted in blue, soft-interacting
    (wounded) nucleons in red, and spectator nucleons in light grey. The reddish tube represents
    the incoming proton, and its transverse size is proportional to the $pN$ total cross section.}
  \label{fig_sketch_2}
\end{figure*}

\begin{figure*}[!hbp]
  \vskip 0.5cm
  \begin{minipage}[h]{0.49\linewidth}
    \center{\includegraphics[width=1\linewidth]{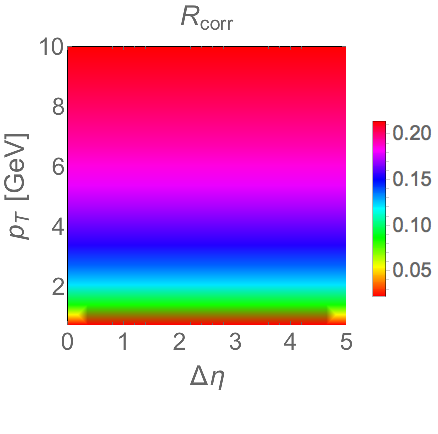} \\ \vskip -0.5cm {\Large(a)}}
  \end{minipage}
  \begin{minipage}[h]{0.49\linewidth}
    \center{\includegraphics[width=1\linewidth]{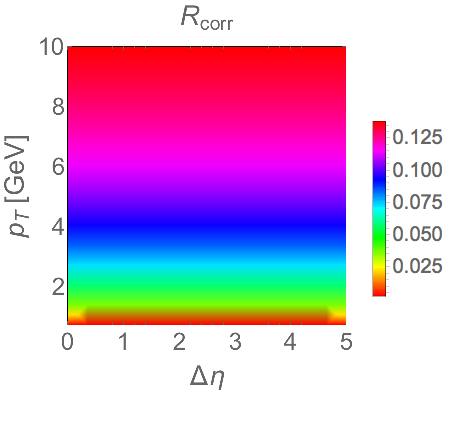} \\  \vskip -0.5cm {\Large(b)}}
  \end{minipage}  
  \caption{Correlation factor as a function of $\Delta \eta$ and \pt for different
    starting points of the QCD evolution, namely $Q_0^2$ = 0.5 GeV$^2$ (a), and $Q_0^2$
    = 1.0 GeV$^2$ (b).}
  \label{rcor}
\end{figure*}

In the optical approximation, which does not include NN correlations and considers the nucleon size much
smaller than the internucleon distance, the cross section of DPS in $pA$ collisions for large A can
be written as follows ~\cite{Strikman:2001gz}:
\begin{equation}
  \sigma_{\mathrm pA}\,=\,A\,\frac{\sigma_1\,\sigma_2}{\sigma_{eff}}\,+\,\sigma_1\,\sigma_2\,\int d^2b\,T^2(b)\,,
  \label{ST}
\end{equation}
where $b$ is the impact parameter of the proton, and $T(b) = \int_\infty^\infty \rho(b, z) dz$ is the standard
nuclear profile function obtained from the nuclear density $\rho(b,z)$. The first term in Eq. (\ref{ST}) is
the contribution of the impulse approximation, in which two partons of the proton interact with two partons
of a single nucleon of the target nucleus (Fig. \ref{fig_sketch_2}a). The second term describes the interaction
of two partons in the proton with two partons of two different nucleons of the nucleus, neglecting parton--parton
correlations in the projectile proton (Fig. \ref{fig_sketch_2}b).
Using a realistic nuclear density, for $A\ge 40$ one can calculate the ratio of the DPS contributions
in \textit{pA} and \textit{pp} scattering as follows ~\cite{Strikman:2001gz}:
\begin{equation}
  r(A)\,=\,\frac{\sigma_{pA}^{DPS}}{\,\sigma_{pp}^{\rm DPS}}\,=\,1\,+\,1.1\,\left(\frac{\sigmaeffp}{\mbox{15 mb}}\right)
  \,\left(\frac{A}{40}\right)^{0.39}\,(1\,+\,R_{\rm corr})\,.
  \label{eq2}
\end{equation}
In Eq. (\ref{eq2}), $R_{\rm corr} = f(x_1,x_2) / f(x_1)f(x_2)-1$ accounts for the longitudinal correlations
of the constituents of the projectile proton due to the pQCD evolution \cite{Blok:2012jr}. In the \textit{pp}
case, correlation effects leads to a decrease of $\sigmaeffp$ by the factor $(1+5 R_{\rm corr})$ as compared
to the uncorrelated (mean field) model. Numerical calculations were performed under the assumption that
the DGPD (Double Generalized Parton Distributions) are factorized at the scale $Q_0^2$ \cite{Blok:2011bu}. 
For different models of double parton correlations at a low resolution scale, see Ref. \cite{Rinaldi:2013vpa}
and references therein. It was found that for large \pt, the  factor $5 R_{\rm corr} = 0.5 \div 1$ allows
to reproduce the measured values of $\sigmaeffp(NN)$; see \cite{Blok:2017alw} and references therein.
For the kinematics we discuss in this work, a typical value is $R_{\rm corr} \sim 0.15$, see Fig. \ref{rcor}. 
Taking $\sigmaeffp$ = 20 mb leads to the expectation that the ratio of DPS to LT contributions is
enhanced in \textit{pPb} collisions by a factor $r(200) \sim 4$. For minijets with \pt of a few GeV,
one expects $\sigmaeffp \sim$ 30~mb, leading to $r(200) \sim 5$. However this enhancement is somewhat
reduced due to the leading twist shadowing effect which requires a detailed modeling of the particular
kinematic domains \cite{Frankfurt:2011cs}, hence this effect will be considered elsewhere.

One can try to observe the predicted enhancement of DPS in $pA$ scattering at the LHC by comparing
\textit{pp} and \textit{pA} data. However this would require comparing two different sets of data
in a somewhat different kinematics. An alternative strategy we suggest in this paper is to explore
the strong dependence of the DPS/LT ratio on the impact parameter of the $pA$ collision. In the mean
field approximation (\textit{cf.} Eq. (\ref{ST})), we can write:
\begin{equation} 
  R_{\rm DPS/LT}(b)\,=\,\sigmaeffp\,T(b)\,, 
  \label{eq4}
\end{equation} 
which corresponds to a very large enhancement of DPS for central $pA$ collisions. We can write 
\begin{equation}
  \frac{d\sigma_{pA}^{\rm DPS}}{d^2b}\,=\,\sigma_{pN}\,T_A(b)\,+\,\sigma_1\,\sigma_2\,T^2(b)\,.
  \label{DPST2}
\end{equation}
In Eq. (\ref{DPST2}),
we removed the superscript (DPS) in the first term on the right hand side to indicate that $\sigma_{pN}$
includes the leading twist contribution to the DPS cross section since it is also linear in $T(b)$.
Hence, Eq. (\ref{DPST2}) gives a model--independent prediction for the $b$--dependence of DPS in terms
of the elementary DPS \textit{pp} cross section, $\sigma_1$ and $\sigma_2$, and of T(b).

Obviously, one cannot fix the impact parameter of the collision, but one can still define centrality
classes, for example using the method adopted by the ATLAS collaboration \cite{Aad:2015zza}. An evidence
of the validity of such a procedure is that it reproduces correctly the rate of jet production in the
kinematics where the parton of the proton carries a moderate Bjorken $x$, like $x \le 0.1$.

To make realistic predictions for the DPS-related observables we perform the calculation in several
steps, extending the existing Monte Carlo generator for the production of dijets
\cite{Alvioli:2014sba,Alvioli:2014eda,Alvioli:2017wou}, which allows to calculate the interaction probability distribution
as a function of the number of wounded nucleons and of the \textit{pA} centrality. We take into account
the finite transverse spread of the parton distribution in nucleons, and correlations between nucleons
in the nucleus \cite{Alvioli:2009ab}.

\section{Inclusive DPS beyond mean field approximation}
\label{sec:sec2}

The generalized double parton distributions necessary for the calculation of the DPS off nuclei were
calculated in Ref. \cite{Blok:2012jr} as sum of two terms, as in Eq. (\ref{eq2}) and as illustrated
in Fig. \ref{fig_sketch_2}.

\begin{figure*}[!htp]
  \centerline{\includegraphics[width=0.8\textwidth]{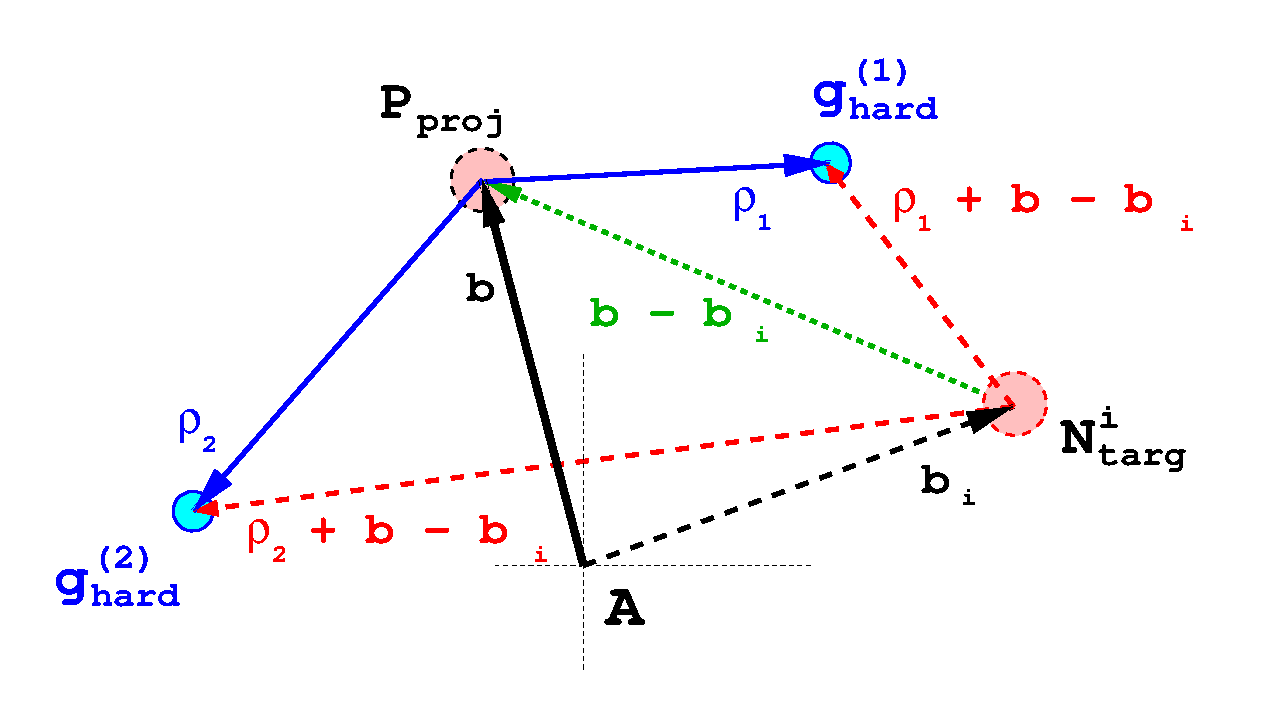}}
  \vskip -0.5cm
  \caption{Sketch of the transverse geometry of the the double parton collisions.
    The incoming proton, $P_{\rm proj}$, is pointed by the vector $b$ form the nucleus'
    center, while the $i$-th nucleon in the target, $N^{i}_{\rm target}$, is pointed by the
    vector $b_i$. In this work, the hard interaction points $g^{(1,2)}_{\rm hard}$, pointed
    by $\rho_1$ and $\rho_2$ from $P_{\rm proj}$, are integrated over the whole transverse
    plane, event-by-event. The remaining vector notations are self-explaining.}
  \label{fig_sketch}
\end{figure*}

The first term in Eq. (\ref{eq2}) term accounts for the scattering off two partons of the same nucleon.
It can be calculated by a convolution of two double nucleon GPDs plus the pQCD induced correlations,
and corresponds to the impulse approximation. The second term in Eq. (\ref{eq2}) corresponds to scattering
of two partons of the projectile off two partons belonging to two different nucleons of the nucleus.
Figure \ref{fig_sketch} shows the notations used for the various quantities used in this work.
Separating the contribution of scattering off the same nucleon is necessary to account in an
economic way for the existence of parton-parton correlations in the nucleons. To calculate
the DPS cross section accounting for a finite transverse spread of the parton distributions we
introduce the quantity $f_N(x, Q^2, \rho)$, describing the transverse distribution of partons
in the nucleon, defined as follows:
\begin{equation}
  f_N(x,Q^2, \rho)\,=\,\frac{g(x, Q^2, \rho)}{p(x, Q^2)}\,,
  \label{eq6}
\end{equation}
where $g(x, Q^2, \rho)$ is the diagonal generalized parton distribution and $p(x, Q^2)$ is the
parton distribution. The $\rho$ dependence of the generalized parton distribution is given by
the Fourier transform of the two gluon form factor of the nucleon, $F_{2g}(t)$, which is determined
from the analysis of $J/\psi$
exclusive  photoproduction \cite{Frankfurt:2010ea}.
For simplicity we will use an exponential parameterization of $F_{2g}(t)=\exp(Bt/2)$, and will not
write explicitly the dependence of $B$ on $x$ and $Q^2$, thus the parton distribution takes the
form:
\begin{equation}
  f_N(x,\rho)\,=\,\frac{1} {2\,\pi\,B}\,\exp( - \rho^2 /2B)\,.
  \label{eq7}
\end{equation}
The value of $B$ in Eq. (\ref{eq7}) can be extracted from the analysis of the exclusive $J/\psi$
photoproduction.

The geometric factor entering to the DPS cross section can be written as 
\begin{eqnarray}
  D^{2 \otimes any2}(b)&=&\int d\Vec{\rho}_1d\Vec{\rho}_2
  \,f_p(\rho_1)\,f_p(\rho_2)\,\psi_A^2(r_t^{(i)},z_i, r_t^{(k)},z_k)\,\times\nonumber\\
  &&\times\,\sum^A_{i=1}\,f_{N}\left(\left|\Vec{\rho}_1+\Vec{b}-\Vec{r}^{(i)}_t\right|\right)
  \,\sum^A_{k=1}\,f_{N}\left(\left|\Vec{\rho}_2+\Vec{b}-\Vec{r}^{(k)}_t\right|\right)\,,
  \label{massi2c}
\end{eqnarray}
\noindent
which includes both interactions with 
 two different nucleons ($2 \otimes 2$) and the
same nucleon ($2 \otimes 1$) of the target nucleus. The geometric factor for the same nucleon
case is given by:
\begin{eqnarray}
  D^{2 \otimes 1}(b)&=&\int d\Vec{\rho}_1d\Vec{\rho}_2 \psi_A^2(r_t^{(i)},z_i)
  \,f_p(\rho_1)\,f_p(\rho_2)\,\times\nonumber\\
  &&\times\,\sum^A_{i=1}\,f_{N}\left(\left|\Vec{\rho}_1+\Vec{b}-\Vec{r}^{(i)}_t\right|\right)
  \,f_{N}\left(\left|\Vec{\rho}_2+\Vec{b}-\Vec{r}^{(i)}_t\right|\right)\,.
  \label{massi2b}
\end{eqnarray}
The factor for the interaction with two different nucleons, which replaces the
$T^2(b)$ factor in the optical approximation, Eq. (\ref{ST}), is simply given by
the difference $D^{2 \otimes any2}(b) - D^{2 \otimes 1}(b)$:
\begin{eqnarray}
  D^{2 \otimes 2}(b)&=&\int d\Vec{\rho}_1d\Vec{\rho}_2 \psi_A^2(r_t^{(i)},z_i, r_t^{(k)},z_k,)
  \,f_p(\rho_1)\,f_p(\rho_2)\,\times\nonumber\\
  &&\times\,\sum^A_{i=1}\,f_{N}\left(\left|\Vec{\rho}_1+\Vec{b}-\Vec{r}^{(i)}_t\right|\right)  
  \,\sum^A_{k\neq i}\,f_{N}\left(\left|\Vec{\rho}_2+\Vec{b}-\Vec{r}^{(k)}_t\right|\right)\,,
  \label{massi2a}
\end{eqnarray}
%
For our numerical studies, we choose $B$ = 3 GeV$^{-2}$, which corresponds to $x \sim 0.01$
for $Q^2 \sim$ a few GeV$^2$. The effective cross section, $\sigmaeffp$ in Eq. (\ref{sigmaeff}),
is expressed through $B$ as $\sigmaeffp= 8\pi B$, leading to $\sigmaeffp$ = 30~mb for
$B$ = 3 GeV$^{-2}$. Smaller values of $\sigmaeffp$ at large virtualities result in this approach
from pQCD induced correlations \cite{Blok:2010ge,Blok:2011bu,Blok:2013bpa}.

The code developed to calculate Eqs. (\ref{massi2c}-{\ref{massi2a}}) thus allows to obtain the
separate contributions due to the DPS with one (Eq. (\ref{massi2a})) and two (Eq. (\ref{massi2b}))
nucleons, both as a function of \textit{pA} centrality and of the number of wounded nucleons.

\begin{figure}[!htp]
  \centerline{\includegraphics[width=0.75\textwidth]{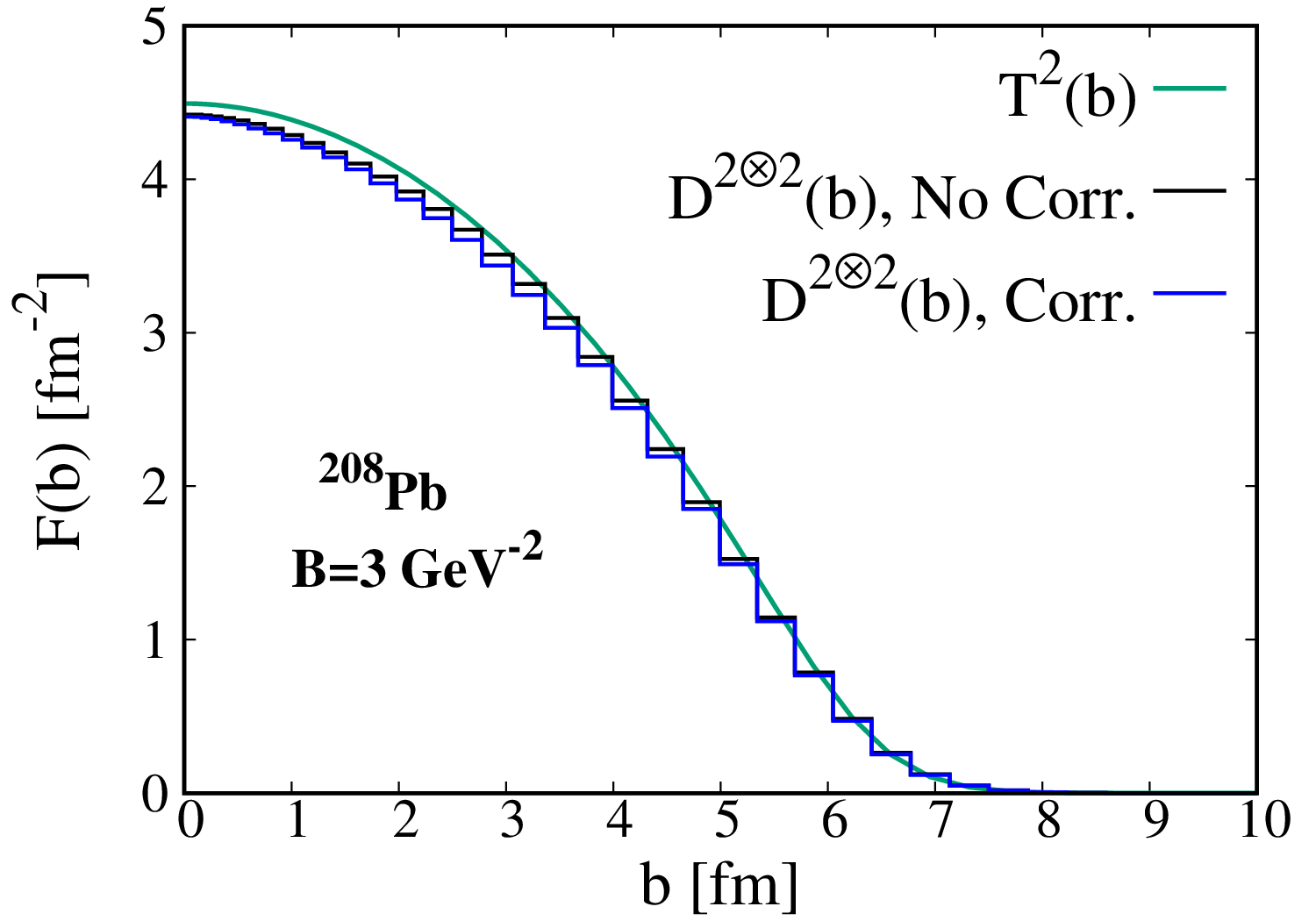}}  
  \caption{The impact parameter distributions for scattering off two nucleons
    in optical approximation, $T^2(b)$, and with finite radius of interaction
    with and without NN correlations, $D^{2 \otimes 2}(b)$, as defined in Eq. (\ref{massi2a}).}
  \label{double22}
\end{figure}

Our numerical results for the $b$--distributions for DPS off two and single nucleon can be compared
with the optical model approximation. In Fig. \ref{double22} we compare $D^{2 \otimes 2}(b)$ and $T^2(b)$.
We find that the $b$--dependent distribution accounting for the finite nucleon size is a bit broader
and the total contribution of the $2 \otimes 2$ term is somewhat smaller than in the optical
approximation. For example, for $pPb$ scattering, $\int d^2b D^{2 \otimes 2}(b) / \int d^2b T^2(b)$ = 0.95,
accounting for finite size, accurate treatment of the surface region of matter distribution (neutron
skin effect, as described in Ref. \cite{Alvioli:2018jls}), and NN correlations. This suppression
factor is close to the correction found in the mean field approximation for the nucleus wave
function accounting for the final nucleon size: $\approx (1 - 2 r_N^2/R_A^2)$ \cite{Blok:2012jr}.

The impulse approximation term $\propto D^{2 \otimes 1}(b)$ obviously does not introduce any corrections
to the cross section integrated over $b$. However, since the elementary cross section corresponds
to the interaction of two nucleons at a finite impact distance, the $b$--distribution of $D^{2 \otimes 1}(b)$
should be somewhat broader than for $T(b)$.

The distribution over $b$ for the leading twist distribution is given by 
\begin{equation}
  S(b)=\int d\Vec{\rho}_1\,f_p(\rho_1)\,\sum^A_{i=1}\,\psi_A^2(r_t^{(i)},z_i )
  \,f_{N}\left(\left|\Vec{\rho}_1+\Vec{b}-\Vec{r}^{(i)}_t\right|\right)\,.
\end{equation}

The difference of S(b) and T(b) is very small, so we do not present the  corresponding plot. 
The double scattering in NN interactions corresponds to a smaller average transverse distance
than a single scattering \cite{Frankfurt:2003td,Blok:2010ge,Blok:2011bu,Blok:2013bpa}. 
So in this case the deviation of the $b$--distribution from $T(b)$
is even smaller. Hence, in the following we will neglect the small difference
between $S(b)$ and $D^{2 \otimes 1}(b)$.

\section{Distribution over the number of wounded nucleons}
\label{sec:sec3}

In order to calculate the distribution over the number of wounded nucleons we need to distinguish
events in which the two interacting partons of the nucleus belong either to the same nucleon or to
two different nucleons. In the first class of events, which is described by $D^{2 \otimes 1}(b)$ in
Eq. (\ref{massi2b}), we need to calculate the distribution over the number of soft interactions
excluding the nucleon involved in the hard interaction. Analogously, we exclude two nucleons in
the case of hard interactions with partons from two different nucleons in the nucleus. The procedure
is a straightforward extension of the one we developed for dijet production \cite{Alvioli:2014sba}.
For each of the two interacting partons of the proton, we assign one particular nucleon as the one
involved in a hard interaction, with probabilities given by:
\begin{equation}
  P_j= {g_N^{(j)}(\rho) \over \sum_{k=1}^A g_N^{(j)}(\rho)}.
  \label{assign}
\end{equation}
Now we need to generate the distribution over number of nucleons involved in soft interactions.  We do it in two ways.
The first approach is based on  the standard Glauber model with an accurate treatment of the distribution of the probability
of the inelastic NN interaction over the relative impact parameter. Another approach includes in addition effects of 
fluctuations of  the strength of interaction of the projectile proton with the target nucleus fluctuates from event to event.
These fluctuations take into account presence of the inelastic diffraction and provide an effective implementation of the high energy Gribov-Glauber picture of hadron - nucleus scattering. We    follow closely the procedure discussed in our paper \cite{Alvioli:2014sba}. We  assign to each incoming proton interaction strength $\sigma$ with probability $P(\sigma)$  - for a detailed discussion see  \cite{Alvioli:2014sba} and calculate averages over a large sample of the events. The variance of the distribution over $\sigma$, $\omega_\sigma= \left< \sigma^2\right>/ \left< \sigma\right>^2 -1$ 
is given by Miettinen - Pumplin relation \cite{Miettinen:1978jb} which expresses $\omega_\sigma $ through the ratio of inelastic and elastic NN cross sections at $t=0$.
For the LHC energies we estimate $\omega_\sigma \approx 0.1$. 
In the analysis of ATLAS  \cite {Aad:2016zif} $R_{pPb}$ was studied as a function of centrality for three values of  $\omega_\sigma=0, 0.11, 0.2$.  
It is remarkable that the best description was found for $\omega_\sigma=0.11$.
 For such $P(\sigma)$  they found $R_{pPb}$ for the charged particle production to be close to  one  for  $\ptp \ge 2\div 3 \,\mbox{GeV}$ and a wide range of rapidities.

Since we count nucleons which were involved in both soft and hard interaction only once, the distribution
for the double nucleon term obviously starts at $\nu=2$, with $\nu$ the total number number of
interacting nucleons.

\begin{figure*}[!htp]
  \centerline{\includegraphics[trim={0 0 0 0 },clip,height=6.0cm]{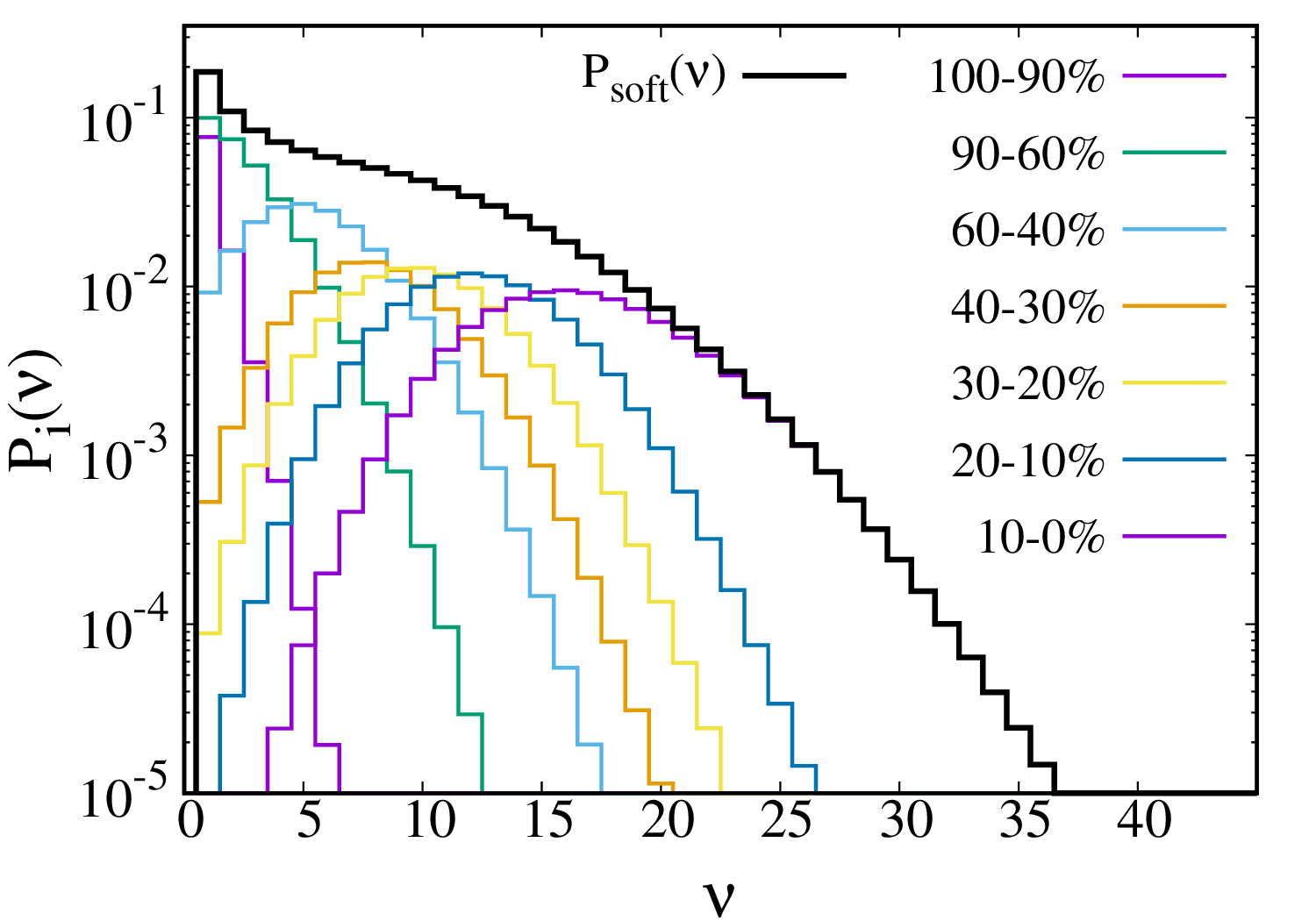}
    \hspace{-0.2cm}\includegraphics[trim={3cm 0 0 0 },clip,height=6.0cm]{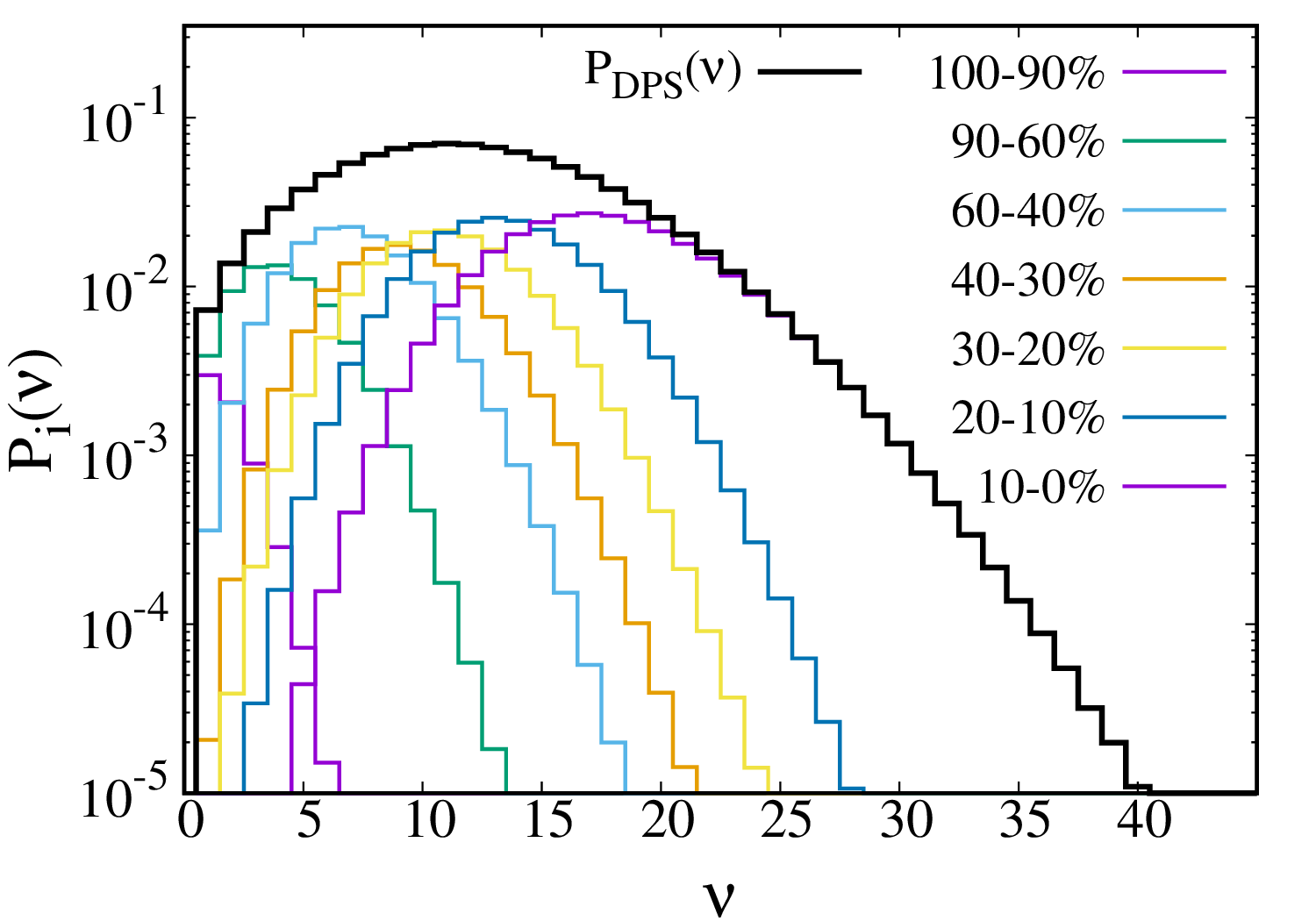}}
  \caption{Left: the centrality distribution of the number of soft collisions.
    Right: the centrality distribution of the number of soft collisions with the double hard trigger.}
  \label{DistrOverNu}
\end{figure*}

The results of the calculation for the distribution over $\nu$ for the no correlation scenario  with account of color fluctuations 
($\sigmaeffp$ = 30 mb) are presented in Fig. \ref{DistrOverNu} for
several centrality classes.  For large $\nu $ the  account of color fluctuations leads to broadening of the distribution over $\nu$.
One can see that for DPS events distribution over $\nu$ is much broader. Parton--parton correlations
lead to an enhancement of the impulse approximation $2 \otimes 1$ term in Eq.~(\ref{massi2b}) by a
factor $\sigmaeffp(m.f.)/ \sigma_{exp}= 1+ 5 R_{\rm corr}$, and of the double nucleon term by a factor
$1+R_{\rm corr}$ \cite{Blok:2012jr}. For the kinematics discussed in this work (presented in Fig.
\ref{fig_sketch}), $R_{\rm corr}\sim 0.15$ (Fig. \ref{rcor}). Hence, its effect for the double scattering
term is pretty small, and we will neglect its residual dependence on impact parameter.
 
To take into account parton--parton correlations in the calculation of the distribution over $\nu$,
it is sufficient to take the impulse approximation term $D^{2 \otimes 1}(b)$ with an additional factor
$(1 + 5R_{\rm corr}) /(1 + R_{\rm corr}) \sim 1.5$ and normalize to the inclusive cross section where
the $D^{2 \otimes 1}(b)$ term is also enhanced by the same factor. 

\section{Transverse energy distribution and extraction of the DPS signal}
\label{sec:sec4}

Let us consider a process in which DPS contributes: for example production of four jets in a special
configuration, or production of two jets and a hadron with a sufficiently large \pt from the underlying
event. The main challenge is that the LT process can also contribute to this special configuration
(\textit{cf.} Fig. \ref{LTvsHT}), leading to the need to rely on a Monte Carlo simulation for a rather
complicated final state.

If we choose a kinematics where soft contributions (including very soft minijets) can be neglected,
there are three contributions to the final state: the leading twist contribution, DPS due to the
interaction with one nucleon and DPS due to the interaction with two nucleons. The first two
contributions are proportional to roughly the number of nucleons along the projectile path.
In the events with a dijet trigger they would result in the same multiplicity of a second dijet
(hadron) for different centralities. At the same time the DPS due to the interaction with two
nucleons should lead to a contribution which grows with centrality much faster (roughly the
square of the number of nucleons along the projectile path). Hence, it is convenient to consider
the ratio of the multiplicity $N$ of the candidate DPS final state (for example dijet plus a pion)
and the multiplicity of the inclusive dijet production in the same kinematics:
\begin{equation}
  N^{D/I}\,=\,N({\rm dijet + pion})/ N({\rm dijet})\,.
\end{equation}
For such a ratio, deviations from linearity in the number of collisions, which were found in
Ref. \cite{Alvioli:2014sba}, practically cancel out. The dependence of $N^{D/I}$ on centrality
is only due to the double nucleon interaction term. 

We follow the procedure developed by ATLAS to define centrality classes \cite{Aad:2015zza}.
They use the transverse energy $-3.2\ge \eta \ge -4.9 $ (i.e. along the nucleus direction)
as a measure of centrality. It was shown in Ref. \cite{Aad:2015ziq} that $\sum E_{\rm T}$ 
in this kinematics is not sensitive to production of hadrons at forward rapidities. 
The distribution over $\sum E_{\rm T}$ as a function of $\nu$ is given in Refs.~\cite{Aad:2015zza,Aad:2015ziq,Aad:2016zif}.

\begin{figure}[!htp]
  \centerline{{\includegraphics[width=0.75\textwidth]{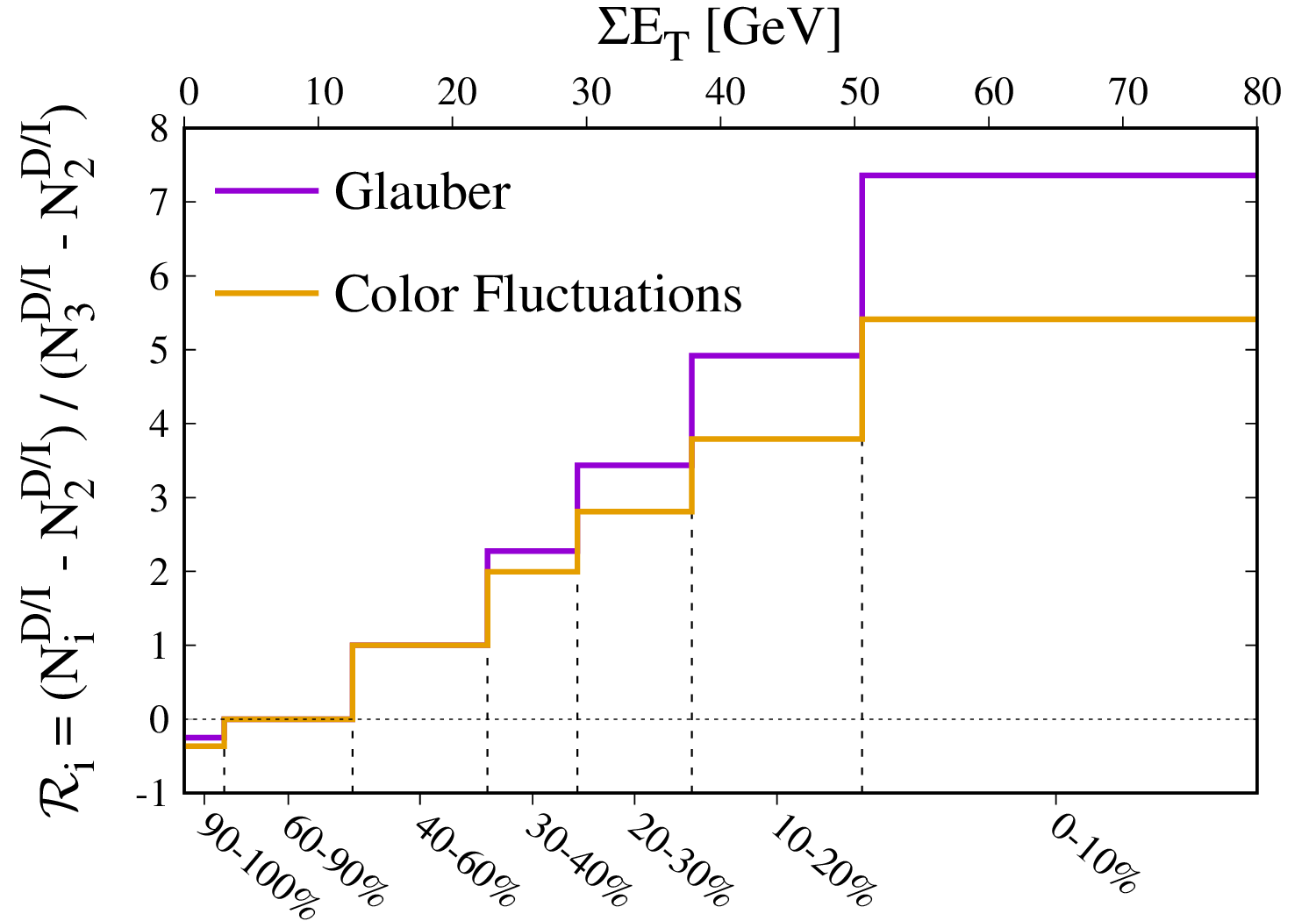}}}
  \caption{Centrality dependence of DPS multiplicity enhancement as a function of $\sum E_{\rm T}$ 
   measured in $-3.2\ge \eta \ge -4.9 $ (along the nucleus direction) which corresponds centrality bins
   denoted in the plot.}
  \label{superrat}
\end{figure}

We define bins in $\sum E_{\rm T}$ as in Refs. \cite{Alvioli:2014eda,Alvioli:2017wou}, and
use the 10\%-20\% (second) bin, in which the first term of Eq.~(\ref{ST})(linear in A) dominates, 
and build the ratio of the differences in multiplicities in the $i^{th}$ centrality bin as follows:
\begin{equation}
\mathcal{R}_{i}\,=\,\frac{N^{D/I}_i - N^{D/I}_2}{N^{D/I}_3 - N^{D/I}_2}\,.
  \label{doubleratio}
\end{equation}
We use for subtraction the second bin since there are significant uncertainties in modeling
the most peripheral bin. In the differences $N^{D/I}_i - N^{D/I}_2$ the contribution of the
terms linear in A cancels out and the $\sum E_{\rm T}$ dependence originates solely from the
geometry of the process. Thus, the dependence of $\mathcal{R}_{i}$ on the momenta of the jets (hadrons)
is expected to be universal. This would provide a crucial test of the overall picture of the
double scattering process. The predicted dependence of $\mathcal{R}_{i}$ on centrality is very strong, as
it is illustrated in Fig. \ref{superrat} for the Color Fluctuation and Glauber models.
One can see that color fluctuations somewhat reduce $\mathcal{R}_{i}$ for most central bin due to additional
smearing over impact parameter. Anyway, the predicted effect is large and should be pretty
straightforward to observe. Note that in our considerations we assumed that both components
of DPS  events originate from the leading twist QCD processes.
So one needs to select the kinematics where for both subprocesses $R_{p Pb}$ is close to one. 
Based on the analysis of  ATLAS \cite{Aad:2016zif} use of the color fluctuation model
with $\omega_\sigma \sim 0.1$ appears to be  preferable. Note also in the kinematics where deviations 
of $R_{pA}$ from one for both subprocesses are small one can estimate related corrections for $\mathcal{R}_{i}$.

An important test of the picture is that $\mathcal{R}_{i}$ should be a universal function of $\sum E_{\rm T}$,
independent on the angle between the dijet and the hadron, and the hadron transverse momentum. 

Let us now consider an example of a process which can be studied using this procedure, the
production of a dijet at forward rapidities, in the range $y$ = 2 $\div$ 4, and a hadron from
an underlying event with a tight cut on the emission angle $\theta = 90^o \pm 10^o$. We performed
the calculations using the \pythia model of the contribution of DPS to the underlying multiplicity.
The results of the calculation were shown in Fig. \ref{LTvsHT}. One can see that in a wide range
of hadron momenta DPS contributes on the scale of 30\% $\div$ 40\% to the \textit{pp} cross section.
Hence, for central $pA$ collisions, the underlying multiplicity should be enhanced by a factor of
at least 2.5 as compared to \textit{pp} collisions. The the underlying multiplicity enhancement 
can be increased by suppressing the LT contribution. For instance, imposing additional requirement 
on the dijet moment imbalance $(p_{\rm T, 1} - p_{\rm T, 2})/(p_{\rm T, 1} + p_{\rm T, 2})<$0.1, 
where $p_{\rm T, 1}$ and $p_{\rm T, 2}$ are the transverse momenta of the leading and subleading
jets respectively, would increase the DPS contribution to 35-50\%.
Also due to a relatively high rate of the discussed
process an accurate subtraction procedure should be possible both for the narrow angle window we discuss,
and for a wider range of the angles. The minimal $\ptp$ of the hadrons for which our calculations are applicable 
follow from the requirement that $R_{pPb}$ should be close to one. Depending on the rapidity of the hadron it 
corresponds to  $ \ptp (hadron) \ge  \mbox{2} \div \mbox{5 GeV}$ \cite{Aad:2016zif}.
Also one has to impose a restriction to the fraction of the momentum  of proton, $x_p$, carried by the parton involved
in the dijet production $x_p \le  0.1$, since for large $x_p$ the centrality dependence gradually changes that maybe
due to selection of smaller size configurations by a large $x_p$ trigger, see discussions in~\cite{Alvioli:2017wou,Alvioli:2014eda}.

A clean separation of the $2 \otimes 2$ contribution would allow to perform a direct measurement
of the parton-parton correlations ($R_{\rm corr}$) (\textit{cf.} Eq. (\ref{eq2})). Knowing the $A^{4/3}$
term it would be possible to measure correlation effects for two partons of the projectile proton
involved in the process (\textit{cf.} Eq. (\ref{eq2})). Also it would make it easier to
extract $\sigmaeffp$ from the linear term. In this case $\sigmaeffp$ is the only parameter which
could be adjusted and it could be determined from the condition that the dependence of the hadron
emission on the azimuthal angle with respect to the dijet should disappear.

\section{Conclusions}

We developed an algorithm for the calculation of the DPS cross section in \textit{pA} scattering
as a function of centrality. We suggested a method to use the centrality to determine the cross
section of DPS due to scattering off two different nucleons. In the long run this would allow to
study parton--parton correlations in nucleons as a function of virtuality and $x$'s. It would be
possible also to look for triple parton scattering \cite{Strikman:2001gz} using a similar strategy.

\section*{Acknowledgments}
\label{Ack}
M.A. acknowledges a CINECA award under ISCRA initiative for making high-performance
computing resources available.
The research of B.B. was supported by Israel Science Foundation under the grant 2025311.
M.S.'s research was supported by the US Department of Energy Office of Science, 
Office of Nuclear Physics under Award No. DE-FG02-93ER40771.
%
%
\bibliographystyle{spphys}
\bibliography{references}
\end{document}